\documentstyle[epsfig,prb,aps,multicol]{revtex}
\begin{document}

\title{Metamagnetism in the 2D Hubbard Model with easy axis}
\author{F. Bagehorn$^1$ and R. E. Hetzel$^2$}
\address{$^1$Institut f\"ur Theoretische Physik,
  Technische Universit\"at Dresden, D-01062 Dresden, Germany\\
  $^2$Max-Planck-Institut f\"ur Physik komplexer Systeme,
  Bayreuther Str.\ 40, D-01187 Dresden, Germany}
\date{\today}
\maketitle

\begin{abstract}
Although the Hubbard model is widely investigated, there
are surprisingly few attempts to study the behavior of such
a model in an external magnetic field. Using the Projector
Quantum Monte Carlo technique, we show that the Hubbard model
with an easy axis exhibits metamagnetic behavior if an external
field is turned on.
For the case of intermediate correlations strength $U$,
we observe a smooth transition from an antiferromagnetic regime
to a paramagnetic phase.
While the staggered magnetization will decrease linearly up to a
critical field $B_c$, uniform magnetization develops only
for fields higher than $B_c$.
\end{abstract}

\draft

\vskip2pc
\begin{multicols}{2}
\section{Introduction}

The Hubbard model is one of the simplest models of strongly correlated
electrons.\cite{Hub63}
The magnetic properties of this model have been extensively studied
for many years.\cite{Scal90,Dag94,Ima92}
But only in a few instances the influence of an external magnetic field
being coupled to the electrons,
has been investigated.\cite{Assa93,Gies93,Voll96}
A very popular approach is the Peierls substitution, i.e.\ a hopping
amplitude of the electrons which depends on the vector potential
of the external field. This is used, e.g., to study the superconducting
properties of a Hubbard ring or torus threaded by a magnetic flux.\cite{Assa93}
It would also be appropriate to calculate Hall coefficients in such
systems.

A different approach is to include a Zeeman term in the Hamiltonian,
i.e.\ to couple the external magnetic field directly to the spins of the
electrons.\cite{Gies93}
This case is well suited for calculating static properties,
such as magnetization.

Since many years it is well known that in alloys with a layered
structure the magnetization shows a specific behavior.
If the planes are itself ferromagnetically ordered but
the coupling between them is antiferromagnetic, one observes that
in an external field the total magnetization first slowly increases
linearly, then suddenly strongly rises
before saturation takes place.\cite{Vogt62}
This was first observed by Becquerel and van den Handel who
coined the term `metamagnetism'.\cite{Becq39}

Especially, since metamagnetic behavior was found in
heavy fermion compounds, the term `meta\-mag\-netism' is used
whenever the magnetic susceptibility $\chi_m(B)$ has a maximum
at a critical field $B_c$, i.e.\ the magnetization $M(B)$ has a
point of inflexion at that field value, even if no phase transition
occurs.

It is widely believed that antiferromagnetic correlations play a crucial
role in meta\-magnetic behavior. Since antiferromagnetic correlations
are inherent in the Hubbard model,
it is a very interesting question to study whether the
Hubbard model shows metamagnetic behavior or not.

Recently Held {\it et al.}\cite{Voll96}\ investigated an anisotropic
Hubbard model in a magnetic field coupled via a Zeeman interaction term.
Using the Grand Canonical Quantum Monte Carlo approach,
they calculated in $d=\infty$ a magnetic phase diagram
and found several phase transitions of first and second order.
It is an open question whether these phases still exist in the
more realistic case of $d=2$ or $3$.

Here, we consider the two-dimensional Hubbard model on a square
lattice in an external magnetic field $B$ coupled to the spins
of the electrons via a Zeeman term. The Hamiltonian $H$ is given by
\begin{eqnarray}
\label{Hamiltonian}
  H & = & \sum\limits_{ij,\sigma}
  t_{ij} c_{i\sigma}^{\dagger}c_{j\sigma}
  + \frac{U}{2} \sum\limits_{i,\sigma} n_{i\sigma}n_{i-\sigma} \nonumber\\
  &   & - \sum\limits_{i} \mu_B B_z s^z_i
\end{eqnarray}
where $t_{ij}$ denotes nearest neighbor hopping,
$B_z$ is the magnetic field parallel to the z axis, and
$s^z_i = \sum_{\sigma} \sigma n_{i\sigma}$ is the spin in z direction.
While the Hamiltonian itself is isotropic, an easy axis along the
z direction will be introduced by the simulational procedure as will
be discussed later on.

\section{Method}
\label{Sec:algorithm}
Here we briefly review the Projector Quantum Monte Carlo (PQMC) method
for fermions in the ground-state.
For a detailed discussion, the reader is referred to Ref.~\onlinecite{Lind92}.

The key idea of the PQMC algorithm is to
project out the ground-state wave function $|\, \Psi_0\rangle$
of a lattice fermion Hamiltonian $H$
from a given trial wave function $|\,\Phi_T\rangle$
by applying the operator $\exp(-\beta H)$
on $|\,\Phi_T \rangle$ according to
\begin{equation}
  \label{projection}
  \lim_{\beta \to \infty}
  \frac{{\rm e}^{-\beta H}\, |\, \Phi_T \rangle}
  {\sqrt{\langle \Phi_T|\, {\rm e}^{-2\beta H}\, |\, \Phi_T \rangle}}
  = |\, \Psi_0 \rangle\, \frac{\langle \Psi_0|\, \Phi_T \rangle}
  {|\langle \Psi_0|\, \Phi_T \rangle|} \;.
\end{equation}
The expectation values of physical quantities $A$ are then obtained from
\begin{equation}
\label{exvalue}
  \langle A \rangle = \lim_{\beta \to \infty}
  \frac{\langle \Phi_{T}|\, {\rm e}^{-\beta H}\, A\, {\rm e}^{-\beta H}\,
  |\, \Phi_{T} \rangle}
  {\langle \Phi_{T}|\, {\rm e}^{-2\beta H}\, |\, \Phi_{T} \rangle} \;.
\end{equation}

Applying the Trotter-Suzuki decomposition\cite{Tro59,Suz76} and the
discrete Hubbard-Stratonovich transformation\cite{Hir83}
to the projection operator, the effect of the projection operator on the
trial state can be rewritten symbolically as a sum over the
Hubbard-Stratonovich spins $\{\sigma\}$,
  ${\rm e}^{-\beta H}\, |\, \Phi_T \rangle =
  \sum_{\{\sigma\}} F(\{\sigma\})\, |\, \Phi_T \rangle$.
The expectation value of a physical quantity $A$ is then obtained from
\begin{equation}
  \label{exspin}
  \langle A \rangle = \frac{\sum\limits_{\{\sigma\},\{\sigma'\}}
  \langle \Phi_T|\, F(\{\sigma\})\, A\, F(\{\sigma'\})\, |\, \Phi_T \rangle}
  {\sum\limits_{\{\sigma\},\{\sigma'\}}
  \langle \Phi_T|\, F(\{\sigma\})\, F(\{\sigma'\})\, |\, \Phi_T \rangle} \;.
\end{equation}

To evaluate these sums, the Monte-Carlo method is used,\cite{Met53} utilizing
\begin{equation}
  |\, \omega(\{\sigma\},\{\sigma'\})\, | =
  |\, \langle \Phi_T|\, F(\{\sigma\})\,
    F(\{\sigma'\})\, |\, \Phi_T \rangle\, |
\end{equation}
as the weight of a configuration of Hubbard-Stratonovich spins.
Since in general $\omega(\{\sigma\},\{\sigma'\})$ can be negative for some spin
configurations $\{\sigma\}$, it can be difficult to evaluate Eq.~(\ref{exspin})
numerically. This problem is often referred to as the minus-sign problem.

All Quantum Monte Carlo simulations suffer from the so-called
`minus-sign' problem though it does not always occur at half-filling.
In the PQMC scheme, the minus-sign problem can be avoided
for the bare Hubbard model at half-filling if one uses a
spin density wave (SDW) ground-state as the trial wave function.
In our simulations, we found that an appropriately chosen SDW
ground-state wave function reduces the minus-sign problem in case of
an additional external magnetic field, too.

The (zero-field) Hubbard model is invariant under SU(2) spin rotation
symmetry.  Since the zero-field Hamiltonian commutes with $S_z$,
the eigenstates of $S_z$ are a natural choice for a basis of states.
At half filling, the ground-state is a state with $S_z=0$.
Therefore, one usually constructs the trial wave function as a direct
product of spin up and spin down wave functions with an equal and
fixed number of electrons in each spin direction. Hence,
the PQMC scheme does not only conserve the total number of electrons,
but moreover it restricts the simulation to states of $S_z=0$.

In order to incorporate the external magnetic field, we have to remove
this constraint. Therefore, we have extended the PQMC algorithm
to all eigenstates of $S_z$.  This is achieved by allowing
the number of electrons with a certain spin to change while still
keeping the total number of electrons fixed.
Hence, we still work in the canonical ensemble appropriate for the
ground-state.
In the framework of the PQMC method, our procedure corresponds
to a manifold of trial wave functions, all differing in spin $S_z$,
which are all sampled by the Monte Carlo method.

To be more specific, let us write a general trial wave function as a sum
over trial wave functions with fixed $S_z$,
\begin{equation}
  \label{trialsum}
  |\, \Phi_T \rangle = \sum_{S_z} \alpha(S_z) |\, \Phi_T(S_z) \rangle \;.
\end{equation}
Now, the expectation value of an operator $A$ which conserves the spin, reads
\end{multicols}
\begin{eqnarray}
  \langle A \rangle &=& \frac{\sum\limits_{\{\sigma\},\{\sigma'\}}
  \sum\limits_{\{S_z,S_z'\}}
  \langle \Phi_T(S_z)|\, \alpha(S_z) F(\{\sigma\})\,
    A\, F(\{\sigma'\}) \alpha(S_z') \, |\, \Phi_T(S_z') \rangle}
  {\sum\limits_{\{\sigma\},\{\sigma'\}}
  \sum\limits_{\{S_z,S_z'\}}
  \langle \Phi_T(S_z)|\, \alpha(S_z) F(\{\sigma\})\,
    F(\{\sigma'\}) \alpha(S_z') \, |\, \Phi_T(S_z') \rangle} \nonumber\\
  &=& \frac{\sum\limits_{\{\sigma\},\{\sigma'\}}
  \sum\limits_{\{S_z\}}
  \frac{\langle \Phi_T(S_z)|\, F(\{\sigma\})\,
    A\, F(\{\sigma'\}) \, |\, \Phi_T(S_z) \rangle}
    {\langle \Phi_T(S_z)|\, F(\{\sigma\})\,
    F(\{\sigma'\}) \, |\, \Phi_T(S_z) \rangle}\,
  \omega(\{\sigma\},\{\sigma'\},S_z)}
  {\sum\limits_{\{\sigma\},\{\sigma'\}}
  \sum\limits_{\{S_z\}}
  \omega(\{\sigma\},\{\sigma'\},S_z)}
\end{eqnarray}
where the absolute value of
\begin{equation}
  \omega(\{\sigma\},\{\sigma'\},S_z) = [\alpha(S_z)]^2
  \langle \Phi_T(S_z)|\, F(\{\sigma\})\,
    F(\{\sigma'\})\, |\, \Phi_T(S_z) \rangle
\end{equation}
\begin{multicols}{2}
\noindent
is now being used as the generalized weight of a configuration.
Application of the Monte-Carlo method is now straightforward.
We want to stress that each point of the configuration space is
still characterized by a definite value of $S_z$.
The original scheme of $S_z=0$ for a half-filled band corresponds
to a choice of
\begin{equation}
  \alpha(S_z)=\left\{
  \begin{array}{c@{\quad}l}
    1 & \mbox{for } S_z=0\\
    0 & \mbox{otherwise} \;.
  \end{array}
  \right.
\end{equation}

Since the Zeeman term is bilinear in the electronic operators
and commutes with the other parts of the Hamiltonian (\ref{Hamiltonian}),
it is easily incorporated into the operator $\exp(-\beta H)$
and the Hubbard-Stratonovich transformation stays unchanged.

The trial wave functions in our scheme are composed of direct
products of Slater determinants of electrons of fixed spin directions,
\begin{equation}
  |\, \Phi_T(S_z) \rangle = |\, \Phi_T^{\uparrow}(S_z) \rangle \otimes
  |\, \Phi_T^{\downarrow}(S_z) \rangle \;.
\end{equation}
Hence no linear combinations of up and of down spins can occur
as would be necessary to construct eigenstates of $S_x$ or $S_y$.
This introduces an easy axis along the z axis into the simulation,
constraining the spins to lie parallel to it.
Since the Hamiltonian conserves spin directions,
the structure of the trial wave function also applies to the projected
ground state wave function, Eq.~(\ref{projection}).
Consequently, the easy axis is conserved throughout the simulation.

\section{Numerical results and discussion}
We have performed simulations for square lattices up to a linear system size
of $L=8$, i.e.\ $N=L\times L$ lattice sites.
On average, we used $m=64$--128 time slices
for our Trotter-Suzuki decomposition.
The number of electrons was set to $N$ and kept fixed throughout the
whole simulation.
As outlined above, however, the number of electrons with a given spin
direction may change during the course of the simulation
so that a net magnetization results.
Typically, we run for an initial warm-up and following measurements
several thousands Monte Carlo sweeps.
This procedure was repeated about 10 times to get independent data
from which the average and the error was computed.

In order to compare our results to the work of Held {\it et al.},
we use $U=2$ (in units of $t$).
We made extensive studies using different projection parameters $\beta$
to ensure proper convergence of the energy and the magnetization, resp.,
to the ground-state.
It can be observed that the energy converges to a $\beta$-independent
value much faster than the magnetization.
However, it turned out that in most cases a value of $\beta=6$ is sufficient
to reach a final value.
Furthermore, we checked that the error due to the Trotter-Suzuki decomposition
is smaller than the statistical error in our data.
This was achieved by varying the number $m$ of time slices.
Then the error due to the decomposition can be estimated from a scaling
of the ground-state energy versus $1/m^2$.

\begin{figure}
\narrowtext
\psfig{file=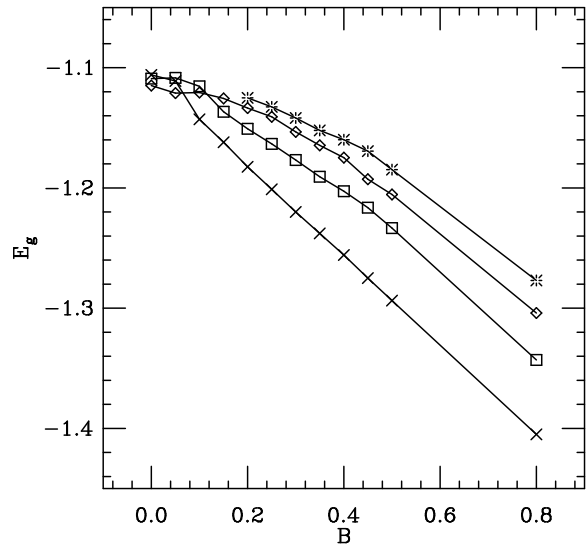,width=\epsfxsize}
\caption{Ground-state energy versus external magnetic field $B_z$.
The symbols correspond to a linear system size of $L=4 (\times)$,
$6 (\Box)$, $8 (\diamond)$, and $\infty(\ast)$.
Lines are guides to the eye only.}
\label{ener1}
\end{figure}

For the trial state $|\,\Phi_T\rangle$, we choose
a spin-density wave state. The external magnetic field is parallel
to the quantization axis of the spins of the electrons
and thus parallel to the easy axis.
Applying such an external field to the system, the electrons can gain energy
by orienting their spin in the direction of the field. Therefore, one would
expect a decrease in the ground-state energy with increasing field.
In Fig.~\ref{ener1} we have depicted the ground-state energy per site,
$E_g$, as measured for three different system sizes.
With increasing magnetic field $B$, the energy is considerably reduced.
For small systems, we observe a linear decrease of $E_g$
while for larger systems the slope slowly changes.

Clearly, there are finite size effects for the energy.
We observe that our data $E_g(B,N)$ scale according to a $1/N$ behavior.
Extrapolating them to $N\to\infty$, we derived $E_g(B)$
for an infinitely large system, too.
For fields of $B=0.2t \ldots 0.8t$, the ground-state energy $E_g(B)$
can nicely be fitted with a quadratic function,
i.e.\ $E_g(B)\propto B^2$.
This quadratic behavior can easily be understood:
For free electrons ($U=0$), it is known that a Zeeman coupling of
the electrons leads to a (temperature independent) van Vleck contribution
to the static susceptibility. Therefore, the energy should show
a quadratic dependence on the magnetic field as confirmed in our
simulations. This behavior, observed for $U=2t$, could be a sign
that the antiferromagnetic correlations introduced by the Coulomb
repulsion $U$, are broken up by the external field $B$.
So we expect some influence on the magnetization as will be discussed below.

In our simulation we computed the spin-spin correlation function
\begin{equation}
S({\bf q}) = \frac{1}{N}\sum_{i,j} {\rm e}^{{\rm i} {\bf q} \cdot
({\bf R}_i - {\bf R}_j)} \langle(n_{i\uparrow}-n_{i\downarrow})
(n_{j\uparrow}-n_{j\downarrow})\rangle \,.
\end{equation}
In order to extrapolate to the thermodynamic limit, we plot $S({\bf q})/N$
vs.\ $1/N$.\cite{Oitm78,Hir85,Hir89}
It should follow a straight line according to
\begin{equation}
S({\bf q}) = N m_q^2 + S_c({\bf q}) \,,
\end{equation}
where $S_c$ is the connected structure factor and $m_q$ the magnetization
\begin{equation}
m_q =  \frac{1}{N}\sum_{i} {\rm e}^{{\rm i} {\bf q} \cdot {\bf R}_i}
\langle(n_{i\uparrow}-n_{i\downarrow})\rangle \,.
\end{equation}
From the extrapolated value $N\rightarrow\infty$, we obtain the square
of the magnetization $m_q$.
We have followed this procedure for $q=0$ and $q=Q\equiv(\pi,\pi)$
to obtain the uniform and the staggered magnetization.

\begin{figure}
\narrowtext
\psfig{file=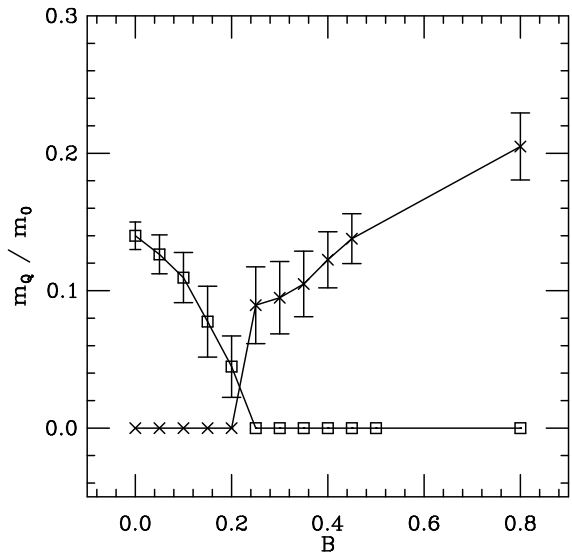,width=\epsfxsize}
\caption{Uniform $(\times)$ resp.\ staggered $(\Box)$ magnetization
versus external magnetic field. Note the transition point from
an antiferromagnetic to a paramagnetic phase at $B_c\approx 0.25 t$.
Lines are guides to the eye only.}
\label{magn1}
\end{figure}

Our results for the magnetizations $m_0(B)$ and $m_Q(B)$
are shown in Fig.~\ref{magn1}.
In zero field, the Hubbard model shows antiferromagnetic order.
With increasing external field, the staggered magnetization clearly
decreases up to a critical field $B_c \approx 0.25t$ where it vanishes.
At about the same field value,
the uniform magnetization strongly rises. The inflexion point is clearly seen
in the uniform magnetization, thus a metamagnetic behavior takes place.

Since in our case the trial wave function is a generalized
spin density wave from which the true ground-state is projected out,
any non-zero staggered magnetization means that the system is still
in an antiferromagnetic ground-state.
This is true up to the critical field $B_c$.

According to a theorem due to Mermin and Wagner, an isotropic two-dimensional
system does not undergo a continuous transition at any finite temperature.
However, long-range order at zero temperature is not excluded.
Although the simulation method cannot deal with $T=0$ directly,
the true ground-state is approached for sufficiently high projection
parameters $\beta$.
Even if we would not have reached high enough $\beta$,
the system would behave effectively as a long-range ordered one if
the correlation length is larger than the system size.

The fact that the system has an easy axis due to the simulational
constraint, is certainly a limitation.  In an anisotropic Heisenberg
model, being the limiting case of large $U$, there exists a minimum field
$B_{\text{tr}}$ at which a so-called spin-flop transition occurs.\cite{White83}
The spins of the electrons will then orient themselves perpendicular to the
external field.  In the isotropic model, this will happen already for an
infinitesimal small field $B_{\text{tr}}$.
Raising the anisotropy, $B_{\text{tr}}$ will increase.\cite{White83}
Further attempts have to be made to clarify whether this scenario
holds for the Hubbard model with intermediate $U$, too.

If one compares our results with those of Held {\it et al.},\cite{Voll96}
there are remarkable differences.
For low temperatures, they found at low external fields a constant,
finite staggered magnetization and vanishing uniform magnetization.
At a critical field of $\tilde{B}_c\approx 0.12 t$, a first order
phase transition takes place leading to a jump in both magnetization curves.
For fields larger than $\tilde{B}_c$, the staggered magnetization remains
zero while the uniform magnetization increases further.
In contrast, our $T=0$ value of $B_c$ is twice as large as $\tilde{B}_c$
and close to the mean-field value of
$B_c^{\text{HF}} \approx 0.27t$.\cite{Voll97}
Besides, we find a rather smooth transition in the staggered magnetization
decreasing steadily up to $B_c$.
Due to strong fluctuations and large statistical errors
close to the phase transition, we were not able to resolve the question
if there is a mixed phase with $m_0 \neq 0$ and $m_Q \neq 0$.
Nevertheless, in our opinion the absence of a jump in $m_Q$
is a strong indication of a second order phase transition
in two dimensions.

\section{Summary}

To summarize, we have studied the half-filled two-dimensional
Hubbard model with an easy axis in an external magnetic field which
was coupled to the spin of the electrons via a Zeeman term.
The model was investigated numerically using an enhanced version
of the Projector Quantum Monte Carlo method.

The model shows in zero field an antiferromagnetic ground-state
which remains present in increasing external magnetic fields up to a critical
field value $B_c\approx 0.25 t$. For higher fields the system is
found to be in a paramagnetic state with field-induced spin orientation.
Our data suggest that the phase transition at $B_c$ should be of
second order.

\acknowledgments
Support from the Deutsche Forschungsgemeinschaft is gratefully acknowledged.
The authors thank K.W. Becker and M.~Vojta for helpful discussions
and valuable comments.


\end{multicols}

\begin{references}
\bibitem{Hub63} J. Hubbard, Proc. R. Soc. London, Ser. A {\bf 276}, 283 (1963);
  {\bf 281}, 401 (1964).
\bibitem{Scal90} D. J. Scalapino, in {\em High Temperature Superconductivity
  Proceedings} (Addison Wesley, 1990), p. 314.
\bibitem{Dag94} E. Dagotto, Rev. Mod. Phys, {\bf 66}, 763 (1994).
\bibitem{Ima92} N. Furukawa and M. Imada,
  J. Phys. Soc. Jpn. {\bf 61}, 3331 (1992).
\bibitem{Assa93} F. F. Assaad, W. Hanke, and D. J. Scalapino,
  Phys. Rev. Lett. {\bf 71}, 1915 (1993).
\bibitem{Gies93} A. Giesekus and U. Brandt,
  Phys. Rev. B {\bf 48}, 10311 (1993).
\bibitem{Voll96} K. Held, M. Ulmke, and D. Vollhardt,
  Mod. Phys. Lett. B {\bf 10}, 203 (1996).
\bibitem{Vogt62} E. Vogt, Z. Angew. Physik {\bf 14}, 177 (1962).
  For a review, see also E. Stryjewski and N. Giordano,
  Adv. Phys. {\bf 26}, 487 (1977).
\bibitem{Becq39} J. Becquerel and J. van den Handel,
  J. Phys. Radium {\bf 10}, 10 (1939).
\bibitem{Lind92} For a recent review, see W. von der Linden,
  Phys. Rep. {\bf 220}, 53 (1992).
\bibitem{Tro59} H. F. Trotter, Prog. Am. Math. Soc. {\bf 10}, 545 (1959).
\bibitem{Suz76} M. Suzuki, Prog. Theor. Phys. {\bf 56}, 1454 (1976).
\bibitem{Hir83} J. E. Hirsch, Phys. Rev. B {\bf 28}, 4059 (1983).
\bibitem{Met53} N. Metropolis, A. W. Rosenbluth, M. N. Rosenbluth, A. Teller,
  and E. Teller, J. Chem.\ Phys. {\bf 21} 1087 (1953).
\bibitem{Oitm78} J. Oitmaa and D. Betts,
  Can. J. Phys. {\bf 56}, 879 (1978).
\bibitem{Hir85} J. E. Hirsch, Phys. Rev. B {\bf 31}, 4403 (1985).
\bibitem{Hir89} J. E. Hirsch and S. Tang, Phys. Rev. Lett. {\bf 62},
  591 (1989).
\bibitem{White83} R. M. White, {\it Quantum Theory of Magnetism}
  (Springer, Berlin, 1983).
\bibitem{Voll97} K. Held, M. Ulmke, N. Bl\"umer, and D. Vollhardt,
  preprint cond-mat/9704209.
\end{references}
\end{document}